# Time domain study of frequency-power correlation in spin-torque oscillators

G. Finocchio,[1] G. Siracusano,[1] V. Tiberkevich,[2] I. N. Krivorotov,[3] L. Torres,[4] B. Azzerboni[1]

[1]Dipartimento di Fisica della Materia e Ingegneria Elettronica, University of Messina, Salita Sperone 31, 98166 Messina, Italy.

[2]Department of Physics, Oakland University, Rochester, Michigan 48309, USA

[3]Department of Physics and Astronomy, University of California, Irvine, 92697-4575 CA, USA.

[4]Departamento de Fisica Aplicada, University of Salamanca, Plaza de la Merced s/n, 37008 Salamanca, Spain.

**Keywords:** exchange biased spin valves, magnetization oscillation, micromagnetic model, spin-torque, nano-oscillators, wavelet.

## Abstract

**This paper describes a numerical experiment, based on full micromagnetic simulations of current-driven magnetization dynamics in nanoscale spin valves, to identify the origins of spectral linewidth broadening in spin torque oscillators. Our numerical results show two qualitatively different regimes of magnetization dynamics at zero temperature: regular (single-mode precessional dynamics) and chaotic. In the regular regime, the dependence of the oscillator integrated power on frequency is linear, and consequently the dynamics is well described by the analytical theory of current-driven magnetization dynamics for moderate amplitudes of oscillations. We observe that for higher oscillator amplitudes, the functional dependence of the oscillator integrated power as a function of frequency is not a single-valued function and can be described numerically via introduction of nonlinear oscillator power. For a range of currents in the regular regime, the oscillator spectral linewidth is a linear function of temperature. In the chaotic regime found at large current values, the linewidth is not described by the analytical theory. In this regime we observe the oscillator linewidth broadening, which originates from sudden jumps of frequency of the oscillator arising from random domain wall nucleation and propagation through the sample. This intermittent behavior is revealed through a wavelet analysis that gives superior description of the frequency jumps compared to several other techniques.**

PACS : 72.25.Ba, 75.75.+a, 75.20.-g

[*]corresponding author. Electronic address: gfinocchio@ingegneria.unime.it

I. INTRODUCTION

Microwave emission driven by spin-polarized current[1] in metallic magnetic nanostructures holds great promise for the development of nanoscale microwave oscillators tunable by magnetic field and current (spin torque oscillators, STO)[2,3,4]. Because the STO frequency is a strong function of the STO output power ($p$), the STO spectral linewidth ($\Delta\omega$) depends on the STO power noise via the non-linear frequency shift, $N$:[5,6,7]

$$\Delta\omega = \Delta\omega_0 \left(1 + (N/\Gamma_{EFF})^2\right) \qquad (1)$$

where $\Delta\omega_0$ and $\Gamma_{EFF}$ are the generation linewidth of a conventional auto-oscillator (that depends on output power and temperature)[6] and the effective non-linear damping, respectively (see Ref. [6] for details). Eq. (1) shows that the stochastic dynamics of strongly non-linear STOs qualitatively differs from the dynamics of classical linear auto-oscillators, in which the linewidth is determined primarily by the phase noise with a negligible contribution from the power noise.

Experimental data of the emitted microwave power as function of the current for the spin-valves described in Ref. [8] are well described by the analytical theory.[7] Furthermore, analysis of a complete set of low temperature experimentally determined parameters of STOs described in Ref.[5] shows that the analytical single-mode theory[6] gives good description of the data for magnetization precession angles up to 70 degrees, where the functional dependence between the output power and the oscillation frequency is approximately linear. However, deviations from the analytical theory are experimentally observed for larger precession angles.[5] In particular, the parametric plots of dimensionless integrated output power ($P_{IOP}$) vs. current ($I$) as well as frequency ($f$) vs. $P_{IOP}$ (see Fig. 3 in Ref. [5]) show that the relationship between these parameters is not a single-valued function at large enough $I$ and $P_{IOP}$. Similar results are also observed for point contact geometries[9] and for low temperature measurements in nanoscale exchange biased spin valves with elliptical (130 nm x 60 nm) 4-nm-thick Py-free layer.[10] In the regime of large-amplitude magnetization dynamics, the measured integrated output power $P_{IOP}$ is not a monotonic function of the current (e.g. see Fig. 7b of Ref. [10] or Fig. 6a of Ref. [9]) while the STO frequency is a monotonically decreasing function of $I$.

In the present work, we numerically study power-frequency correlations for an STO operating in the regime of large precession angles. In our simulations, we identify two qualitatively different regimes of magnetization dynamics at zero temperature: regular (single-mode precessional dynamics) and chaotic. In the regular regime, we find that the dependence of the integrated output power $P_{IOP}$ (the power measured experimentally as the integral of the power spectrum) on the oscillation frequency can be quantitatively predicted from the dependence of the STO non-linear power[11,12] ($P_{NL}$, characterized by non-linear dependence of the power on amplitude) on the oscillation frequency. Our results suggest that the deviations of the predictions of the analytical theory predictions from the experimental data[5] for large precession angles can be explained if $P_{NL}$ is used for data analysis instead of the "linear" power $P_L$. For a range of current densities in the regular regime, our simulations predict a linear dependence of the linewidth on temperature for a fixed current density as expected from Eq. (1) in the low temperature regime (linewidth sufficiently small).[13]

In the chaotic regime observed at large current values, we find that well-defined power-frequency correlation observed in the regular regime of oscillations is lost. Our wavelet-based analysis shows that the chaotic magnetization dynamics in this regime is characterized by sudden jumps of frequency versus time corresponding to micromagnetic events such as domain wall nucleation and propagation through the free layer of the spin valve. Similar frequency jumps have been observed recently in MgO-based magnetic tunnel junction STOs.[14] We show that the large phase noise in this regime of chaotic dynamics can be estimated by integrating the wavelet-instantaneous frequency of the dominant excited STO mode.

## II. NUMERICAL SIMULATIONS

We performed numerical experiments by means of micromagnetic simulations for an STO system similar to that experimentally studied in Ref. [10]. This system is exchange biased spin valve IrMn(8)/Py(4)/Cu(8)/Py(4) (the thicknesses are in nm; Py=$Ni_{80}Fe_{20}$) of elliptical shape (130nm x 60nm). The exchange biased Py layer acts as the pinned layer while the other Py layer is the free layer. We introduce a Cartesian coordinate system where the x-axis and y-axis are the long and the short axes of the ellipse. We simulate, by solving the Landau-Lifshitz-Gilbert-Slonczweski equation, the entire spin-valve with the account of feedback effect of the spin-torque from the free layer on the pinned layer.[15] The effective field consists of the exchange, the magnetostatic and the Oersted fields, and exchange bias field acting on the pinned layer. We use Slonczweski form of spin torque[16] $\mathbf{T}(\mathbf{m_p},\mathbf{m_f})$ (where $\mathbf{m_p}$ and $\mathbf{m_f}$ are the normalized pinned and free layer magnetizations, respectively) for both the pinned and free layers:

$$\mathbf{T}(\mathbf{m_p},\mathbf{m_f}) = \begin{pmatrix} \mathbf{T}_{free} \\ \mathbf{T}_{pinned} \end{pmatrix} = \frac{g}{|e|\gamma_0} \frac{|\mu_B|}{M_s^2} \begin{cases} \dfrac{j}{L_F} \varepsilon(\mathbf{m_f},\mathbf{m_p}) \; \mathbf{m_f} \times (\mathbf{m_f} \times \mathbf{m_p}) \\ \dfrac{-j}{L_P} \varepsilon(\mathbf{m_p},\mathbf{m_f}) \; \mathbf{m_p} \times (\mathbf{m_p} \times \mathbf{m_f}) \end{cases} \quad (2)$$

where $g$ is the gyromagnetic factor, $\mu_B$ is the Bohr magneton, $e$ is the electron charge, $j$ is the current density (positive for current flowing flows from the pinned to the free layer), $L_F$ and $L_P$ are the thicknesses of the free and pinned layer respectively, and $\varepsilon(\mathbf{m_p},\mathbf{m_f})$ is the polarization function given by:

$$\varepsilon(\mathbf{m_p},\mathbf{m_f}) = 0.5 P \Lambda^2 / \left(1 + \Lambda^2 + (1-\Lambda^2)\mathbf{m_p} \bullet \mathbf{m_f}\right) \qquad (3)$$

where $P$ is the current polarization and $\Lambda^2$ is related to the asymmetric giant-magneto-resistance (GMR) parameter $\chi$ as $\Lambda^2 = 1 + \chi$.[16] The magneto-resistance signal is computed as $r(\mathbf{m_p},\mathbf{m_f}) = \frac{1}{N_f} \sum_{i=1...N_f} r_i(\mathbf{m}_{i,\mathbf{p}},\mathbf{m}_{i,\mathbf{f}})$, where $r_i(\mathbf{m}_{i,\mathbf{p}},\mathbf{m}_{i,\mathbf{f}})$ is the magneto-resistance signal of the $i^{th}$ computational cell of the free layer computed with respect to the $i^{th}$ computational cell of the pinned layer $r_i(\mathbf{m}_{i,\mathbf{p}},\mathbf{m}_{i,\mathbf{f}}) = [1-\cos^2(\theta_i/2)]/[1+\chi \cos^2(\theta_i/2)]$ ($\cos(\theta_i) = \mathbf{m}_{i,\mathbf{p}} \bullet \mathbf{m}_{i,\mathbf{f}}$ and $N_f$ is the number of computational cell of the free layer).[16,17]

We make simulations for a fixed external field (68 mT applied at -45 degrees with respect to the long axis of the ellipse) for a range of bias currents. In these simulations we use saturation magnetization of Py 650 $10^3$ A/m, Py exchange constant 1.3 $10^{-11}$ J/m, Gilbert damping parameter $\alpha_F = 0.025$ for the free layer and $\alpha_P = 0.2$ for the pinned layer,[17] the exchange bias field acting on the pinned layer is 70 mT (applied along +45 degrees with respect to the easy axis of the ellipse), $P$ and $\chi$ are 0.38 and 1.5, respectively. The computational cell is 5x5x4 nm$^3$ and the integration time step is 0.2 ps.

For the finite temperature simulations, the thermal fluctuations have been taken into account via adding a stochastic field to the effective field of each computational cell of the pinned and free layers.[18,19,20]

### III. RESULTS

*A) Regular regime*

The magnetization dynamics for this system is excited in a broad range of currents with a critical current density for the onset of auto-oscillations $J_C$=1.2 $10^8$ A/cm$^2$. For zero temperature simulations and for the bias current densities from $J$=1.2 $10^8$ A/cm$^2$ to $J$=2.4 $10^8$ A/cm$^2$, a regular regime where the magnetization dynamics exhibits a single-mode character is observed. Fig.1(a) shows an example of the power spectrum in the regular regime computed for $J$=1.7 $10^8$ A/cm$^2$ by means of the micromagnetic spectral mapping technique (MSMT).[21,22] In this regime, the finite generation linewidth at zero temperature is caused solely by the finite computational time. The regular dynamics is characterized by a spatially uniform coherent precession of the magnetization as illustrated in the inset of Fig. 1(a). Figure 1(b) displays the micromagnetic and the experimental (from Ref. [10]) data of frequency versus current density. The oscillation frequency red shift as a function of current is in qualitative agreement with the analytical non-linear theory.[11,23] Figure 1(b) also shows normalized powers ($P_{IOP}$, $P_L$, and $P_{NL}$) as functions of the current density. The integrated output power $P_{IOP}$ corresponds to the power measured in the experiments and it is computed by integrating the power spectrum:

$$P_{\text{IOP}} = \frac{1}{N_f} \int_0^{f_S/2} \sum_{i=1...N_f} |R_i(f)|^2 df \qquad (4)$$

where $R_i(f)$ is the Fourier transform of $r_i(\mathbf{m}_{i,\mathbf{p}}, \mathbf{m}_{i,\mathbf{f}})$ and $f_S$ is the sampling frequency.

For a single excited mode, the "linear" power $P_L$ is defined as the square of the amplitude of the resistance oscillations of the giant-magneto-resistance signal $r(\mathbf{m_p}, \mathbf{m_f})$.[11] From the numerical point of view, we use the following approximation $r(\mathbf{m_p}, \mathbf{m_f}) = a_0 + a_1 \sin(2\pi f_A t + \varphi_1) + a_2 \sin(4\pi f_A t + \varphi_2)$ ($f_A$ is the frequency of the excited mode) to compute $P_L = a_1^2 + a_2^2$.

The nonlinear power $P_{NL}$ of the excited spin wave mode is computed as follows: in the regular regime, the trajectory $\gamma$ of the magnetization vector is closed for every computational cell (see Fig. 2(a) for an example) and, therefore, the curve $\gamma$ splits the surface of the unit sphere into two parts (the smaller $S_1$ and the larger $S_2$). The power $P_{NL}$ is defined as the sum of the $S_1$ for all the computational cells of the free ($N_f$) and the pinned ($N_p$) layers of the STO:

$$P_{\text{NL}} = \frac{1}{4\pi N_f} \sum_{i=1...N_f} \iint_{S_{1i}} dS + \frac{1}{4\pi N_p} \sum_{i=1...N_p} \iint_{S_{1i}} dS \qquad (5)$$

Defined in such a way, the power $P_{NL}$ is proportional to the canonical action corresponding to the closed loop $\gamma$ and represents proper generalization of the oscillation power for the case of spatially-nonuniform and non-circular magnetization precession.

The power $P_{IOP}$ is computed in the frequency domain, while $P_L$, and the $P_{NL}$ are computed in the time domain. The "red shift" functional dependence of the oscillation frequency on current density (see Fig. 1(a)) for the regular region given by our simulations is in agreement with the experimental data.[10] The calculated frequencies of magnetization self-oscillations are smaller than the ferromagnetic resonance frequency, $f_{FMR}$, estimated to be 6.7GHz by means of spin torque ferromagnetic resonance calculations.[24,25] Simulations of frequency as a function of power give negative non-linear frequency shift $N/(2\pi) \approx -1.45$GHz computed near the critical current density as:[7]

$$N = 2\pi(f_{J_C} - f_{FMR})/P_L(J_C) \tag{6}$$

The parametric dependence of $P_{NL}$ and $P_L$ on the oscillation frequency is shown in Fig. 1(c). This dependence is non-monotonic for $P_{NL}$ with the maximum of $P_{NL}$ achieved around the current density $J_{max}=1.8 \cdot 10^8$ A/cm$^2$. Fig. 2(b) shows the trajectories (x-z plane) of the magnetization of one computational cell (the trend is independent of the computational cell) for $J$=1.5, 1.8, 2.2 $10^8$ A/cm$^2$. As the current density increases, the magnetization trajectories first expand up to $1.8 \cdot 10^8$ A/cm$^2$, and then become deformed and intersections with curves at lower current appear. In this current density range the $P_{NL}$ decreases. In contrast, $P_L$ is monotonic in frequency. It shows behaviour qualitatively similar to that of $P_{NL}$ for small current densities $J < J_{max}$, but approaches saturation for $J > J_{max}$. Fig. 1(b) clearly demonstrates that for large values of current, $P_{NL}$ has very similar functional dependence on current as the experimentally measurable quantity, $P_{IOP}$. In contrast, the dependence of $P_L$ on current is qualitatively different from that of $P_{IOP}$. This means that one can not use $P_L$ for calculation of such parameters as nonlinear frequency shift for current densities far above the supercritical current density.[5] Our numerical results show the functional dependence of $P_{NL}$ and $P_{IOP}$ on oscillation frequency can be well approximated by an empirical function – third-order polynomial:

$$P_{IOP} = a_3 f^3 + a_2 f^2 + a_1 f + a_0 \tag{7}$$

(see for example the solid line in Fig. 1(c) for $P_{NL}(f)$ fitting). By considering the current density range where the $P_{IOP}$ and $P_N$ are proportional to each other ($J = 1.3$-$2.0 \cdot 10^8$ A/cm$^2$), the polynomial coefficients computed for $P_{IOP}$ are proportional to the polynomial coefficients of $P_{NL}$ (i.e. $a_{3,IOP}/a_{3,NL} = a_{2,IOP}/a_{2,NL} = a_{1,IOP}/a_{1,NL} = a_{0,IOP}/a_{0,NL}$).

*B) Chaotic regime*

For current densities larger than $2.4 \cdot 10^8$ A/cm$^2$ we observe zero-temperature broadening of the linewidth (chaotic regime), as evident from the magnetoresistance power spectrum computed via the MSMT in Fig. 3(a) for the current density of 3.2 $10^8$ A/cm$^2$. This spectral broadening is related to the loss of temporal coherence due to the presence of non-uniformities in the spatio-temporal distribution of the magnetization.[26,27] In contrast to the regular regime, our numerical results show a lack of correlation between power and frequency in the chaotic regime.

A time-frequency characterization of the magnetorestance signal for large currents gives the possibility to identify the origin of the linewidth broadening at zero temperature. We systematically study the micromagnetic wavelet scalogram (MWS) [28] of the magnetoresistance signal. We use the complex Morlet function as the wavelet mother:

$$\psi_{u,s} = \frac{1}{\sqrt{s\pi f_B}} e^{j2\pi f_c\left(\frac{t-u}{s}\right)} e^{-\left(\frac{t-u}{s}\right)^2/f_B} \tag{8}$$

which has the best time($\sigma_t$)-frequency($\sigma_f$) resolution because of its Gaussian envelope ($\sigma_f = \left(2\pi s\sqrt{f_B}\right)^{-1}$, $\sigma_t = 0.5s\sqrt{f_B}$).[29] The variables $u$ and $s$ are the translation and scale parameters, $f_B$ and $f_C$ are the characteristic parameters. For the complex Morlet wavelet function with $f_C = 1$, it is possible to relate the Fourier frequency directly to the scale $f = \frac{f_S}{s}$, where $f_S$ is the sampling frequency.[30]

Fig. 3(b) shows the MWS (white/black color corresponds to the largest/smallest wavelet amplitude) computed for $J=3.2\ 10^8$ A/cm$^2$ ($f_B=300$, $f_C=1$). This figure shows that there are two sources of the linewidth broadening: (i) continuous modulation of the frequency of the dominant excited mode and (ii) discontinuous frequency jumps (see points A and B in Fig. 3(b)). By analyzing the time evolution of the spatial distribution of the magnetization close the points A and B, we observe nucleation of a domain wall at one side of the device and its subsequent propagation to the other side (see supplementary movie in Ref.31).

Another advantage of the wavelet analysis is the possibility to estimate the phase noise $\phi(t)$ directly from the time domain data through integration of the instantaneous wavelet frequency $f_i(t)$ of the dominant exited mode (the frequency that has the largest value of the wavelet transform at a given moment of time) $\phi(t) = 2\pi \int_0^t f_i(\tau)d\tau + \phi(0)$. The instantaneous phase and frequency can also be estimated in a simple way from the zero crossing of the voltage time traces as described in Ref. [32]: if $f_0$ is the oscillation frequency the instantaneous phase, for the time ($t_i$) where the zero crossing occurs, can be computed as $\phi(t_i) = n\pi - 2\pi f_0 t_i$ ($n$ is even (odd) for crossing with a positive (negative) slope). Fig. 3(c) shows a comparison of the instantaneous frequency $f_i$ computed with the wavelet analysis (blue circles) and the voltage crossing (solid line). The wavelet-based analysis gives the range of instantaneous frequency fluctuations that is in a better agreement with the linewidth of the power spectrum in Fig. 3(a) (the linewidth computed with a Lorentzian fit is 278MHz).

*C) Effect of the thermal fluctuations*

In the presence of thermal fluctuations, we calculate the $P_{IOP}$ power as a function of temperature. We found that the functional dependence of the $P_{IOP}$ on the oscillation frequency is qualitatively independent of temperature as displayed in Fig. 4(a) (temperatures from 50K to 300K at step of 50K, an offset is used for clarity) for the current values in the regular regime. We also calculated the temperature dependence of the linewidth (the simulation time was $1\mu s$ which corresponds to a frequency resolution of 1MHz); the results are summarized in Fig. 4(b) for three current densities $J$=1.2, 1.8, and 2.2 $10^8$ A/cm$^2$. In agreement with the analytical theory[33] and experimental data (e.g. see Fig. 4(d) in Ref.[5]), we find broad spectra near the threshold current with non-linear dependence of the linewidth on temperature (e.g. $J$=1.2 $10^8$ A/cm$^2$) corresponding to a GMR-signal with high noise level. For higher current densities, the linewidth decreases (Fig.4(b), $J$=1.8 and 2.2 $10^8$ A/cm$^2$) by approximately one order of magnitude and the dependence of the linewidth on temperature is approximately linear at a fixed current density value. In this current region, the signal-to-noise ratio is significantly improved. Our results are in qualitative agreement with analytical calculations as predicted by Eq. (1) in the low temperature regime[13] and experimental measurements.[8]

IV. CONCLUSIONS

In conclusion, we performed a numerical experiment to identify the origin of the discrepancy of analytical theory and experimental data observed in some experimental studies of large-amplitude STO dynamics. We found two qualitatively different regimes of magnetization dynamics at zero temperature: regular (single-mode precessional dynamics) and chaotic. In the regular regime for moderate amplitudes of oscillations, the functional dependence of the integrated output power (the experimentally measured quantity) on the oscillation frequency is linear, and it is well described by the analytical theory[11]. Our studies reveal that micromagnetically calculated nonlinear power $P_{NL}$ gives a better description of the integrated output power for large amplitudes of magnetization oscillations of the regular regime than the linear power $P_L$ used in the current analytical theories. We also found the linewidth as a function of temperature is a linear function in a range of currents larger than the threshold current. These findings are important for modeling of STOs working in a wide range of current densities. In contrast, the chaotic regime is not described by the single-mode analytical theory. We find that in the chaotic regime, the oscillator linewidth broadening originates from sudden jumps of the oscillator frequency arising from random domain wall nucleation and propagation across the STO free layer. This intermittent behaviour is revealed through a wavelet analysis that gives superior description of the frequency jumps compared to other techniques.

ACKNOWLEDGMENTS

This work was supported by Spanish Project under Contracts No. MAT2008-04706/NAN and No. SA025A08. I. N. K. gratefully acknowledges the support of DARPA, NSF (grants DMR-0748810 and ECCS-0701458) and the Nanoelectronics Research Initiative through the Western Institute of Nanoelectronics.

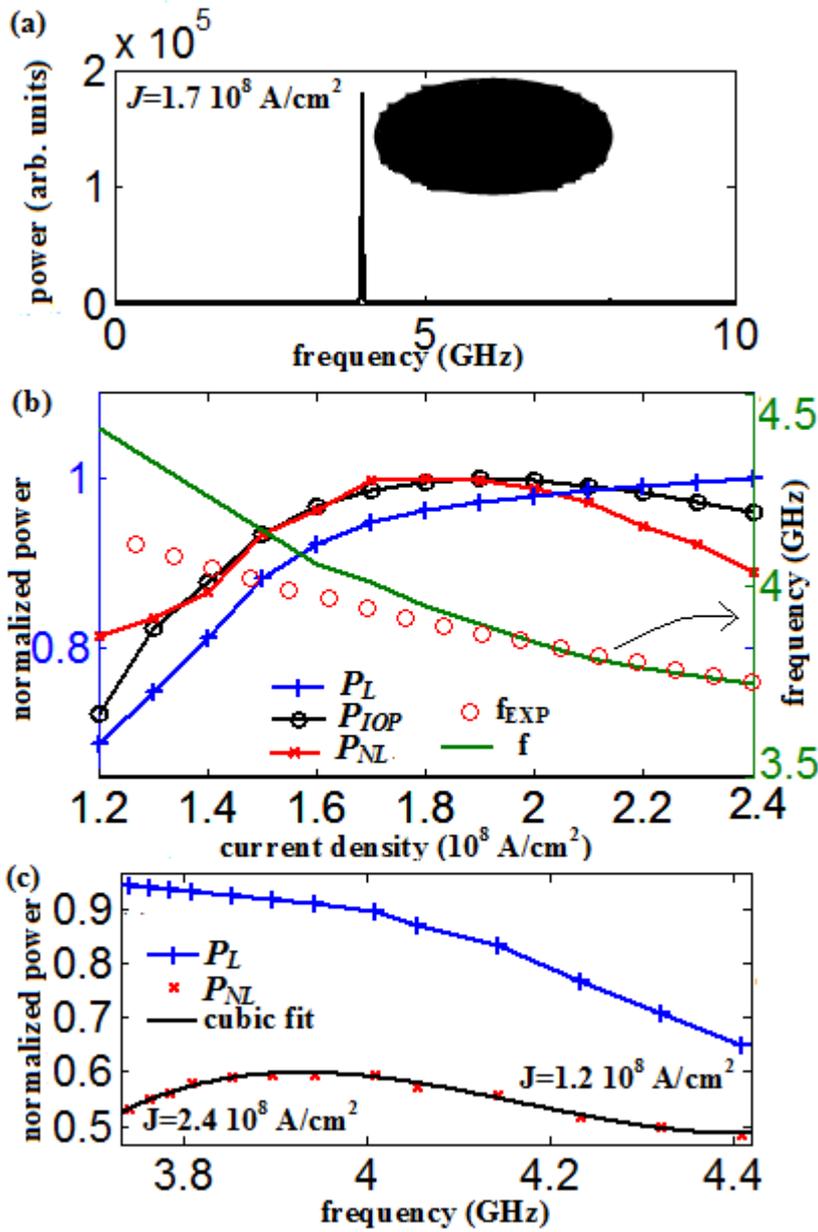

Figure 1. (color online) (a) Example of Fourier spectrum computed by means of the micromagnetic spectral mapping technique in the regular dynamics regime for a current density $1.7 \times 10^8$ A/cm$^2$. Inset: spatial distribution of the dominant excited mode. (b) Micromagnetic (solid line) and experimental from reference [10] (circle "o") frequency of the dominant excited mode, linear power ($P_L$), integrated output power ($P_{IOP}$), and non-linear power ($P_{NL}$) as function of the current density. (c) Parametric dependence of the non-linear power and linear power on the generated frequency and a cubic fit of the non-linear power.

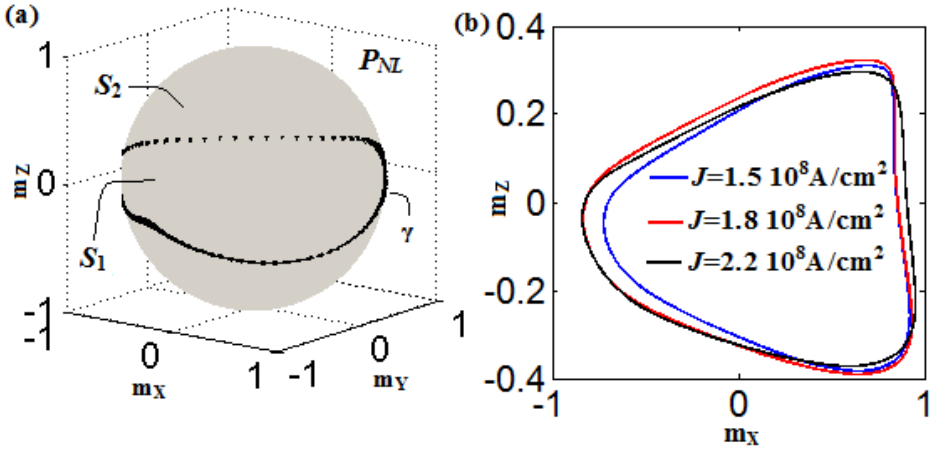

Figure 2 (color online) (a) Example of the calculated magnetization trajectory and illustration of the non-linear power definition. (b) Example of the trajectories of the magnetization projected in the x-z plane) of one computational cell (the trend is independent of the computational cell) for $J$=1.5, 1.8, 2.2 $10^8$ A/cm$^2$.

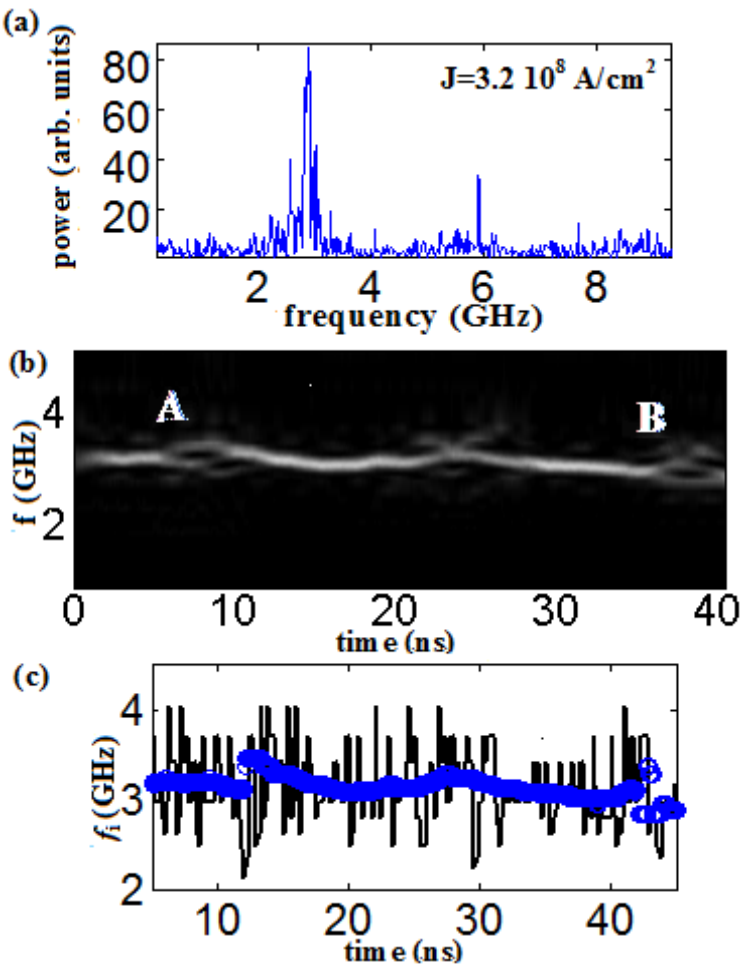

Figure 3: (color online) (a) and (b) Fourier spectrum computed by means of the micromagnetic spectral mapping technique and micromagnetic wavelet scalogram (white/black color corresponds to the largest/smallest wavelet amplitude) for a current density 3.2 $10^8$ A/cm$^2$ respectively. (c) Comparison between instantaneous frequency computed via the wavelet transform (blue circle) and the technique developed in Ref [30] (black solid line).

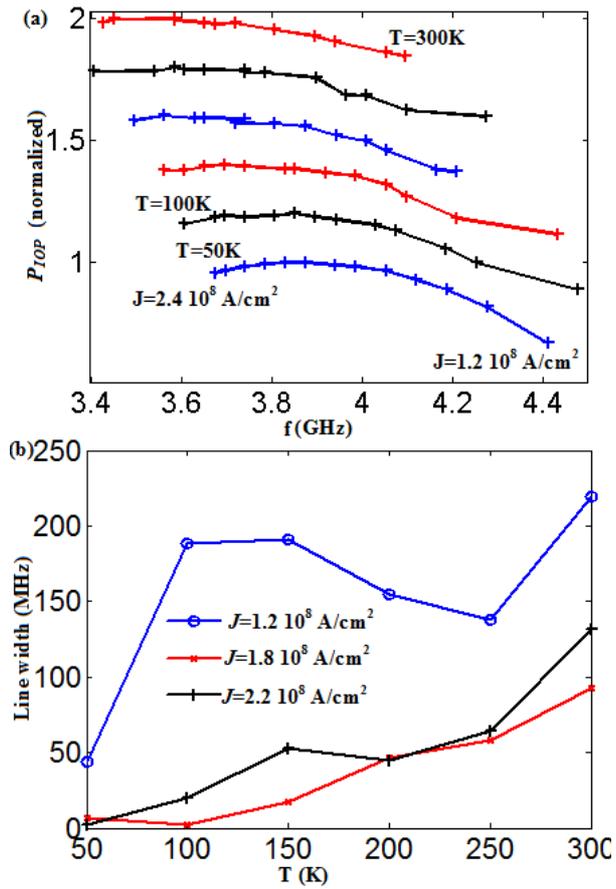

Figure 4: (color online) (a) Temperature dependence of integrated output power on oscillation frequency of the dominant excited mode (an offset is applied for each curve). (b) Temperature dependence of the linewidth for three current density values $J$=1.2, 1.8, and 2.2 $10^8$ A/cm$^2$).